# Design of reversible low-field magnetocaloric effect at room temperature in hexagonal MnMX ferromagnets

Jun Liu[1,2,*], Yurong You[2], Ivan Batashev[1], Yuanyuan Gong[2], Xinmin You[1], Bowei Huang[1], Fengqi Zhang[1], Xuefei Miao[2], Feng Xu[2,**], Niels van Dijk[1], Ekkes Brück[1]

[1]*Fundamental Aspects of Materials and Energy (FAME), Faculty of Applied Sciences, Delft University of Technology, Mekelweg 15, 2629 JB Delft, The Netherlands*

[2] *MIIT Key Laboratory of Advanced Metallic and Intermetallic Materials Technology, School of Materials Science and Engineering, Nanjing University of Science and Technology, Nanjing 210094, People's Republic of China*

[*]Corresponding author. E-mail: liujun@njust.edu.cn, Tel: ++86-25-84303411

[**]Corresponding author. E-mail: xufeng@njust.edu.cn, Tel: +86-25-84303411

**Abstract**

Giant magnetocaloric effect is widely achieved in hexagonal MnMX-based (M = Co or Ni, X = Si or Ge) ferromagnets at their first-order magnetostructural transition. However, the thermal hysteresis and the low sensitivity of the magnetostructural transition to the magnetic field inevitably lead to a sizeable irreversibility of the low-field magnetocaloric effect. In this work, we show an alternative way to realize a reversible low-field magnetocaloric effect in MnMX-based alloys by taking advantage of the second-order phase transition. With introducing Cu into Co in MnCoGe alloy, the martensitic transition is stabilized at high temperature, while the Curie temperature of the orthorhombic phase is reduced to room temperature. As a result, a second-order magnetic transition with negligible thermal hysteresis and a large magnetization change can be observed, enabling a large reversible magnetocaloric effect. By both calorimetric and direct measurements, a reversible adiabatic temperature change of about 1 K is obtained under a field change of 0-1 T at 304 K, which is larger than that obtained in a first-order magnetostructural transition. To get a better insight into the origin of these experimental results, first-principles calculations are carried out to characterize the chemical bonds and the magnetic exchange interaction. Our work provides a new understanding of the MnCoGe alloy and indicates a feasible route to improve the reversibility of the low-field magnetocaloric effect in the MnMX system.

# 1. Introduction

A strong coupling between a first-order structural change and a magnetic phase transition, the so-called magnetostructural transition (MST), can bring about a giant magnetocaloric effect (MCE) [1-4]. Magnetic refrigeration (MR) based on the MCE is environmentally friendly and highly energy-efficient, and has therefore been considered to be a promising technology to replace the conventional gas refrigeration [5,6]. As one of the candidate materials for MR at room temperature, the hexagonal MnMX-based (M = Co or Ni, X = Si or Ge) ferromagnets, which experience a martensitic transition, attract a lot of attention owing to its significant advantages: (i) a strong magnetostructural coupling can easily be established and highly be tuned between the Curie temperatures of two phases by elements substitution [7-9], the introduction of vacancies [10,11] and hydrostatic pressure [12,13]; (ii) the featured paramagnetic-ferromagnetic (PM-FM) type MST gives rise to the same sign of the enthalpy change during the martensitic and the magnetic transition, leading to a higher magnetic entropy change ($|\Delta S_m|$) than other magnetocaloric materials [8,14,15]; (iii) the compounds with desired compositions can be produced easily.

Nevertheless, the first two points in turn bring about evident disadvantages into MnMX-based system. Firstly, the first-order nature of the MST inevitably results in the occurrence of thermal/magnetic hysteresis. Secondly, according to the Clausius-Clapeyron equation, the giant entropy change greatly decreases the sensitivity of $T_t$ to the magnetic stimulus [12,16]. Both drawbacks result in a significant functional fatigue of the giant MCE in magnetic cycles [17]. Our previous work showed that a giant reversible $|\Delta S_m|$ (>20 Jkg$^{-1}$K$^{-1}$) under the field change of 0-5 T can be obtained by minimizing the thermal hysteresis in Mn$_{0.9}$Fe$_{0.2}$Ni$_{0.9}$Ge$_{1-x}$Si$_x$ system [18]. While this sizeable reversibility of MCE would significantly degrade for a low field variation [18]. In addition, Liu *et al.* measured the adiabatic temperature change ($\Delta T_{ad}$) during the MST in MnCo$_{0.95}$Ge$_{0.97}$ that the reversible value is only about 0.7 K under a field change of 1.9 T [19]. Constructing an active magnetic refrigerator is desirable to operate under a cyclic low magnetic field change of 1 T at ambient temperature [20,21]. Under this cyclic field, the magnetic field-induced MST is irreversible and the associated MCE is reduced, which directly hinders the potential for application of the MnMX system as magnetic coolant. Thus, it is of key importance to realize a reversible low-field (0-1 T) MCE at room temperature in MnMX alloys.

Considering the intrinsic disadvantages of MST, we alternatively focus on the second-order phase transition (SOMT) in MnMX alloy. Despite the moderate MCE of the second-order transition, its transition is continuous, which may provide a feasible alternative to optimize the MnMX system. Fig. 1 shows the main design schematic. Here, we choose MnCoGe as the host material due to its large saturation magnetic moment (4.13 $\mu_B/f.u.$) and magnetic transition near room temperature [22,23]. In order to realize our aim, two criteria should be met: (i) the martensitic transition needs to remain in the high temperature range to lower its impact on the magnetic transition; (ii) the Curie temperature of orthorhombic phase ($T_C^o$) with a higher magnetization should be tuned to room temperature. In this work, we introduce Cu on the Co site of stoichiometric MnCoGe to fulfill these goals. For an increasing Cu content, the martensitic transition temperature varies slightly and lies in the PM region. Simultaneously, $T_C^o$ is reduced to room temperature. Indirect (calorimetric) and direct measurements are adopted to evaluate the MCE performance of the MnCo$_{1-x}$Cu$_x$Ge system in which the tunable, reversible and sizeable values of $\Delta T_{ad}$ and $\Delta S_m$ are achieved. In addition, the obtained experimental results are compared with first-principles calculations to unravel the physical mechanism that controls the MCE.

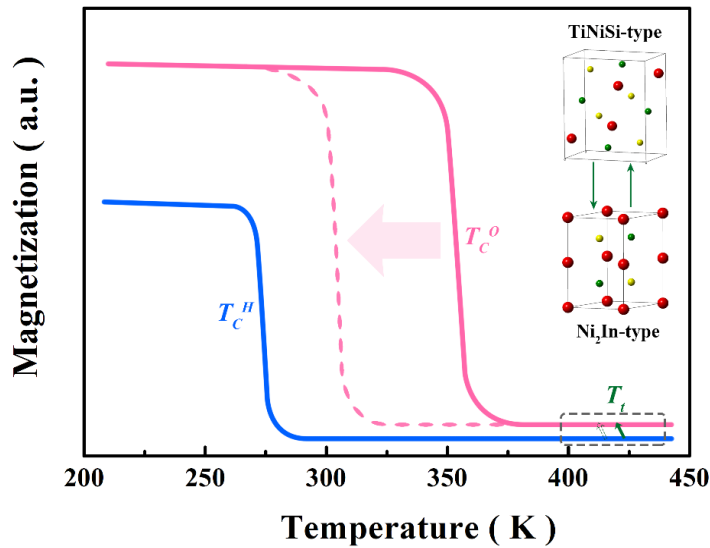

Fig. 1. Schematic of the alloy design for a room-temperature SOMT. Solid pink and blue lines are magnetization curves of the orthorhombic and hexagonal phases in the MnCoGe alloy, respectively. The solid green arrow represents the martensitic transition. As indicated by the dashed pink line and dashed green arrow, $T_C^o$ is aimed to be reduced to room temperature. The structural transition is stabilized in the high-temperature PM region (grey dashed box). The Ni$_2$In-type and TiNiSi-type crystalline structures are shown.

## 2. Experimental details and Calculational method

### 2.1. Sample preparation and characterization

Polycrystalline samples with nominal compositions MnCo$_{1-x}$Cu$_x$Ge ($x$ = 0.00, 0.06, 0.07, 0.08) were prepared by arc-melting high-purity raw materials for four times to ensure the homogeneity. Then ingots sealed in vacuum quartz tubes were annealed at 1123 K for 5 days and slowly cooled to room temperature in 18 h. The thermal properties were measured in a differential scanning calorimeter (DSC, TA Instrument Q2000) with a heating/cooling rate of 10 K/min. The magnetic properties were measured on a superconducting quantum interference device (SQUID, Quantum Design MPMS 5XL) with the reciprocating sample option mode. The crystal structures were characterized by powder X-ray diffractometer at room temperature (XRD, PANalytical X'Pert PRO). The structural parameters were refined using the Fullprof package [24]. The calorimetric measurements in applied magnetic field were carried in a home-built DSC with a Pieter-cell (details described in Ref. [25]), from which the $\Delta S_m$ and $\Delta T_{ad}$ can be calculated. A direct $\Delta T_{ad}$ measurement device was employed to measure the $\Delta T_{ad}$ under cyclic fields of 1.1 T at a rate of 1.1 T/s, as described in Ref. [26]. Here, the powder sample was compressed into a capsule and then a thermocouple was buried to guarantee a good thermal contact.

### 2.2. Density functional theory

Electronic localized function (ELF) calculations on the basis of density function theory was performed using the Vienna *ab initio* Simulation Package (VASP) [27]. We implemented Perdew-Burke-Ernzerhof (PBE) pseudopotentials with generalized gradient approximation (GGA) exchange correlation functions. A plane-wave cutoff energy of 500 eV and 9×9×9 **k** points were chosen. Here, a supercell of 8-unit cells with a hexagonal lattice structure were considered. The geometry optimizations for the lattice parameters and atomic site occupancies were performed on the reported experimental lattice parameters of MnCoGe [28]. Additionally, the interatomic exchange interaction calculations were performed using the Green's function Korringa-Kohn-Rostoker formalism (SPR-KKR) [29]. The potential was treated within the atomic sphere approximation (ASA). The lattice parameters of orthorhombic phases were linearly interpolated from the experimental values of MnCoGe and MnCo$_{0.92}$Cu$_{0.08}$Ge, as shown in Table I.

## 3. Results

### 3.1. Structural information

Fig. 2 shows the XRD patterns of MnCo$_{1-x}$Cu$_x$Ge ($x$ = 0, 0.06, 0.07 and 0.08) alloys at room temperature. All samples crystallize in the TiNiSi-type orthorhombic structures (*Pnma*, space group 62) which indicates that the structural transition occurs above room temperature. A small amount of the hexagonal structure less than 4% can be also obtained which may result from the effect of residual stress during the grinding process [30]. According to the site occupation principle in MnMX alloys [31], Cu atom would occupy the *4c* site of Co atom. The lattice parameters and the site occupation determined from the Rietveld refinement are listed in Table 1. Evidently, an increase in lattice parameter *a* and *c* can be observed, while parameter *b* shrinks with the introduction of Cu. Consequently, the unit cell volume expands from 160.89 Å$^3$ ($x$ = 0) to 162.14 Å$^3$ ($x$ = 0.08) due to the larger atomic radius of Cu atom compared to Co.

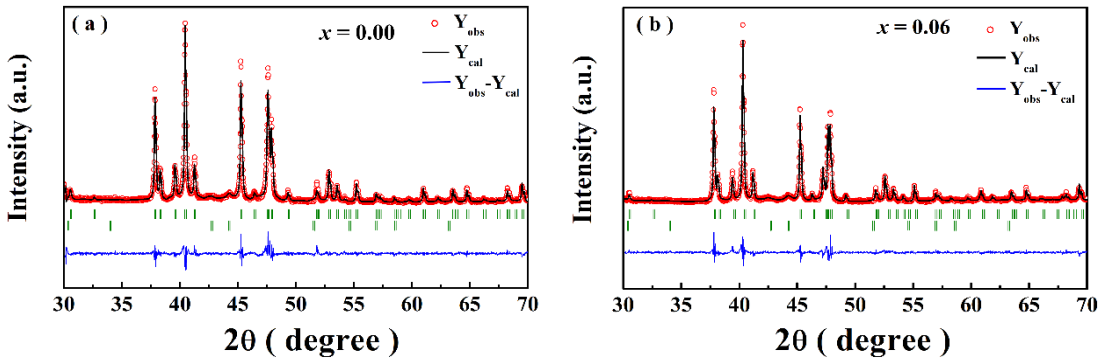

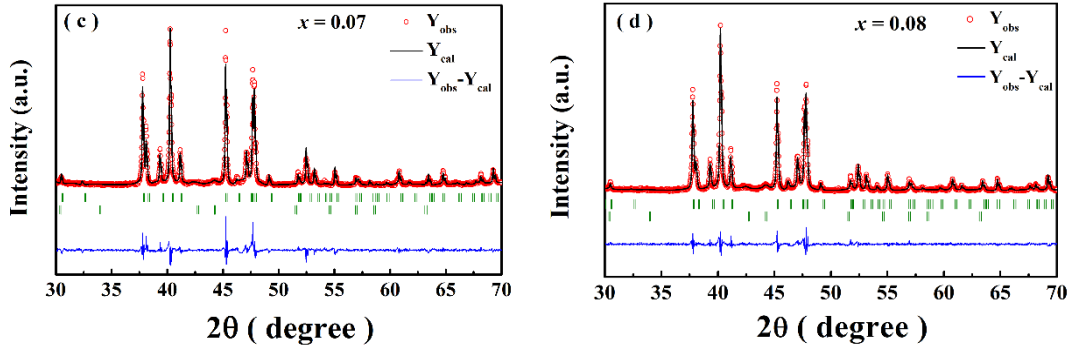

Fig. 2. XRD patterns at room temperature with the corresponding Rietveld refinements for MnCo$_{1-x}$Cu$_x$Ge: (a) $x = 0.00$, (b) $x = 0.06$, (c) $x = 0.07$ and (d) $x = 0.08$.

Table 1. Lattice parameters, unit cell volume, site occupation, goodness of the Rietveld refinement ($R_{wp}$), Curie temperature of the orthorhombic phase and the adiabatic temperature change under field change of 0-1 T for MnCo$_{1-x}$Cu$_x$Ge ($x = 0.00, 0.06, 0.07$ and $0.08$). All atoms occupy Wyckoff position *4c* ($x$, 1/4, $z$) in the orthorhombic phase. Due to the restriction of measurement temperature range, the $\Delta T_{ad}$ for the sample with $x = 0.00$ cannot be obtained.

| Sample | $x = 0.00$ | $x = 0.06$ | $x = 0.07$ | $x = 0.08$ |
|---|---|---|---|---|
| $a$ (Å) | 5.9643(2) | 5.9993(1) | 6.0131(2) | 6.0195(1) |
| $b$ (Å) | 3.8211(1) | 3.8153(1) | 3.8146(1) | 3.8136(1) |
| $c$ (Å) | 7.0597(2) | 7.0607(1) | 7.0638(2) | 7.0634(2) |
| $V$ (Å$^3$) | 160.89(1) | 161.61(1) | 162.03(1) | 162.14(1) |
| $x_{Mn}$ | 0.0301(7) | 0.0259(7) | 0.0414(8) | 0.0225(7) |
| $z_{Mn}$ | 0.6898(5) | 0.6848(4) | 0.6905(5) | 0.6865(4) |
| $x_{Co/Cu}$ | 0.1562(7) | 0.1594(6) | 0.1528(8) | 0.1565(6) |
| $z_{Co/Cu}$ | 0.0569(5) | 0.0594(4) | 0.0590(6) | 0.0561(5) |
| $x_{Ge}$ | 0.2702(5) | 0.2685(5) | 0.2657(6) | 0.2660(5) |
| $z_{Ge}$ | 0.3817(4) | 0.3748(3) | 0.3821(5) | 0.3770(3) |
| $R_{wp}$ (%) | 2.99 | 2.95 | 4.09 | 2.81 |
| $T_C^O$ (K) | 348.9 | 314.0 | 305.9 | 295.9 |
| $\Delta T_{ad}$ (K) | - | 1.1 | 1.0 | 1.0 |

## 3.2. Thermal and magnetic properties

To investigate the structural transition temperature ($T_t$), DSC curves of MnCo$_{1-x}$Cu$_x$Ge ($x$ = 0.00, 0.06, 0.07, 0.08) alloys are shown in Fig. 3(a). The large exothermic/endothermic peaks with an obvious thermal hysteresis ($\Delta T_{hys}$) correspond to the martensitic/austenitic transitions in MnMX-based alloys. The temperature marked as $M_s$, $M_f$, $A_s$ and $A_f$ denote the start and finish points of martensitic transition and austenitic transition, respectively. With increasing Cu content, $T_t$ (defined as $T_t = (M_s+M_f)/2$ or $(A_s+A_f)/2$) firstly increases and then decreases. For the sample with $x \leq 0.08$, $T_t$ remains in a relatively high temperature range above room temperature, which suggests the first-order structural transition is insensitive to the introduction of Cu into the Co site. In Fig. 3b, the thermomagnetic (*M-T*) curves clearly show the typical continuous PM-FM-type magnetic transition with negligible hysteresis below $T_t$. In comparison, the magnetic transition temperature $T_C^O$ decreases monotonously with increasing Cu content, and can be tuned to room temperature, as shown in Fig. 3c.

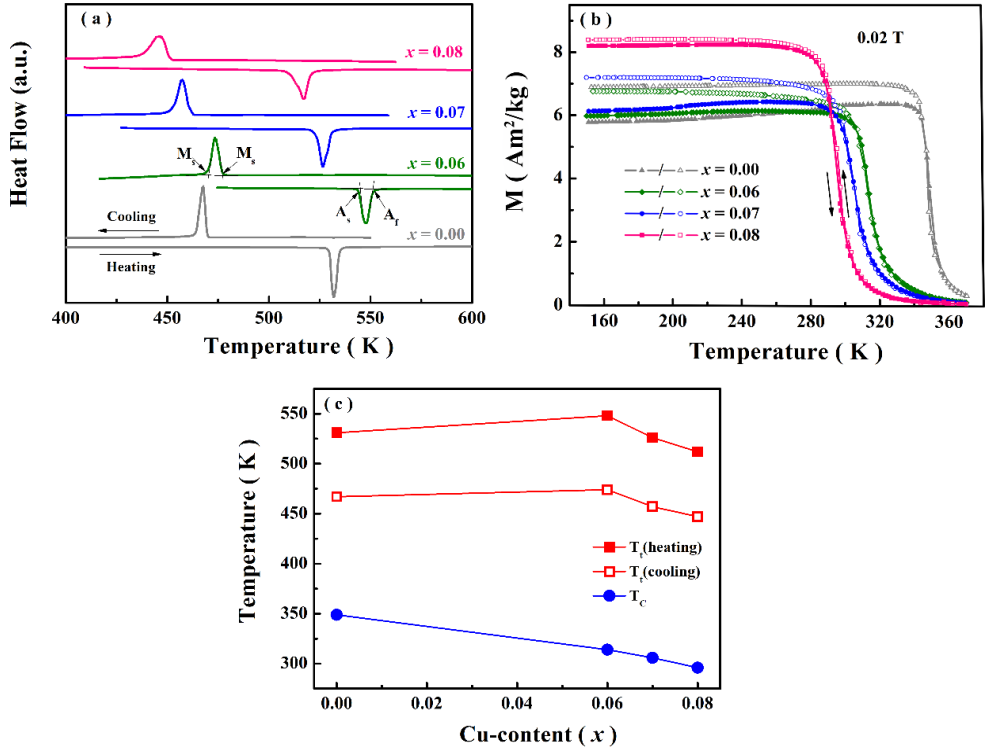

Fig. 3. (a) DSC curves and (b) *M-T* curves in a field of 0.02 T in the heating and cooling processes for MnCo$_{1-x}$Cu$_x$Ge alloys. (c) Evolution of the characteristic temperatures- $T_t$(heating), $T_t$(cooling) and $T_C^o$ as a function of the Cu content.

It is generally believed that the saturation magnetization ($M_s$) of the ferromagnetic phase is important to obtain a large MCE [32]. As displayed in Fig. 4a, the magnetization (*M-B*) curves at 5 K demonstrate that samples with $x \leq 0.08$, which show an orthorhombic structure have a large $M_s$ with values above 110 Am$^2$/kg (3.75 $\mu_B$/f.u.). This is because the structural distortion during the martensitic transition brings about a larger Mn-Mn separation in the orthorhombic structure, which leads to a narrower 3*d* band widths and a larger exchange splitting between the majority and minority bands [33]. A slight reduction in $M_s$ is attributed to the substitution of magnetic Co by non-magnetic Cu. Therefore, benefited from the above optimization, a considerable magnetization change during the room-temperature magnetic transition under 1 T is achieved, as shown in Fig. 4b.

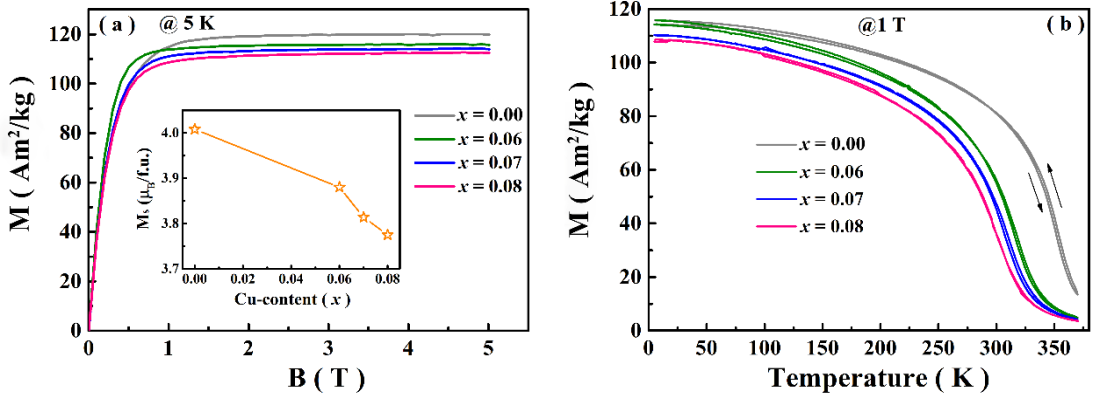

Fig. 4. (a) Magnetization curves at 5 K for MnCo$_{1-x}$Cu$_x$Ge alloys. The inset shows the saturation magnetization as a function of *x*. (b) *M-T* curves in a field of 1 T for MnCo$_{1-x}$Cu$_x$Ge alloys.

*3.3 Magnetocaloric performance*

The most straightforward assessment of MCE, $\Delta T_{ad}$ as a function of temperature is derived from the calorimetric measurements in magnetic field. As shown in Fig. 5a, a large $\Delta T_{ad}$ of 1.1, 1.0 and 1.0 K under a field change of 0-1 T can be achieved for the sample with $x$ = 0.06, 0.07 and 0.08, respectively. Due to the negligible thermal hysteresis during the transition, the obtained value of $\Delta T_{ad}$ is expected to be reversible. Direct $\Delta T_{ad}$ measurements are also performed for the sample with $x$ = 0.07. During the measurement, the temperature sweeping mode with cyclic magnetic fields is adopted. As shown in Fig. 5b, the largest $\Delta T_{ad}$ of about 1.1 K is achieved for $\Delta B$ = 1.1 T, which is reversible during the field oscillations. Although lower than the values obtained for the famous $(Mn,Fe)_2(P,Si)$, $La(Fe,Si)_{13}$-based alloys and pure Gd, the measured values of $\Delta T_{ad}$ are larger than the reversible values in the MnMX family for a first-order MST under the same field change [19,34], and are comparable with those in the Ni-Mn-based Heusler alloys, $Mn_2Sb$-based alloys, $(Hf,Ta)Fe_2$-based alloys and $Tb_x(Dy_{0.5}Ho_{0.5})_{1-x}Co_2$ compounds at the first-order magnetic transition [35-38]. Additionally, the smooth magnetic transition results in a wider working temperature range for the MCE of about 30 K, as estimated by the full width at the half maximum in $\Delta T_{ad}$ versus temperature presented in Fig. 5b.

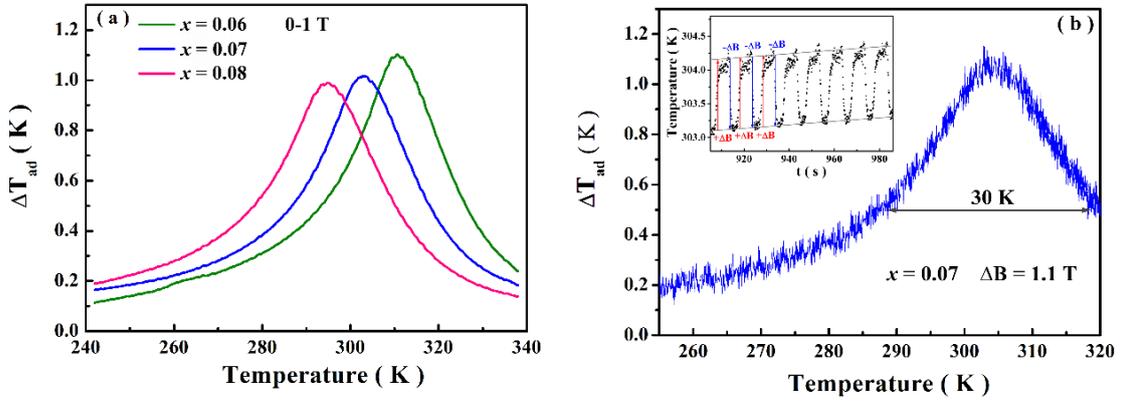

Fig. 5 (a) Temperature dependence of $\Delta T_{ad}$ for a field change of 0-1 T in $MnCo_{1-x}Cu_xGe$ ($x$ = 0.06, 0.07 and 0.08) alloys. (b) Direct measurement of $\Delta T_{ad}$ in the temperature sweeping mode for the sample with $x$ = 0.07 for $\Delta B$ = 1.1 T. The full width at the half maximum is also listed. The inset shows the data for the temperature versus time signal around $T_C^O$ during the direct $\Delta T_{ad}$ measurement.

Besides $\Delta T_{ad}$, another important parameter $\Delta S_m$ is also derived from the calorimetric measurements in magnetic field upon cooling. The maximum value is about 1.6, 1.5 and 1.5 $Jkg^{-1}K^{-1}$ for sample $x$ = 0.06, 0.07 and 0.08, respectively, as shown in Fig. 6a. To evaluate the reversibility of this entropy change, direct measurements of the entropy change for a field change of 1 T is shown in Fig. 6b. The exothermic (endothermic) peak on applying (removing) represents the conventional MCE. During 100 cycles in this study, the peak height is almost unchanged, which demonstrates that the phase transitions possess a good reversibility and stability. The calculated value is consistent with the indirect value (not shown here).

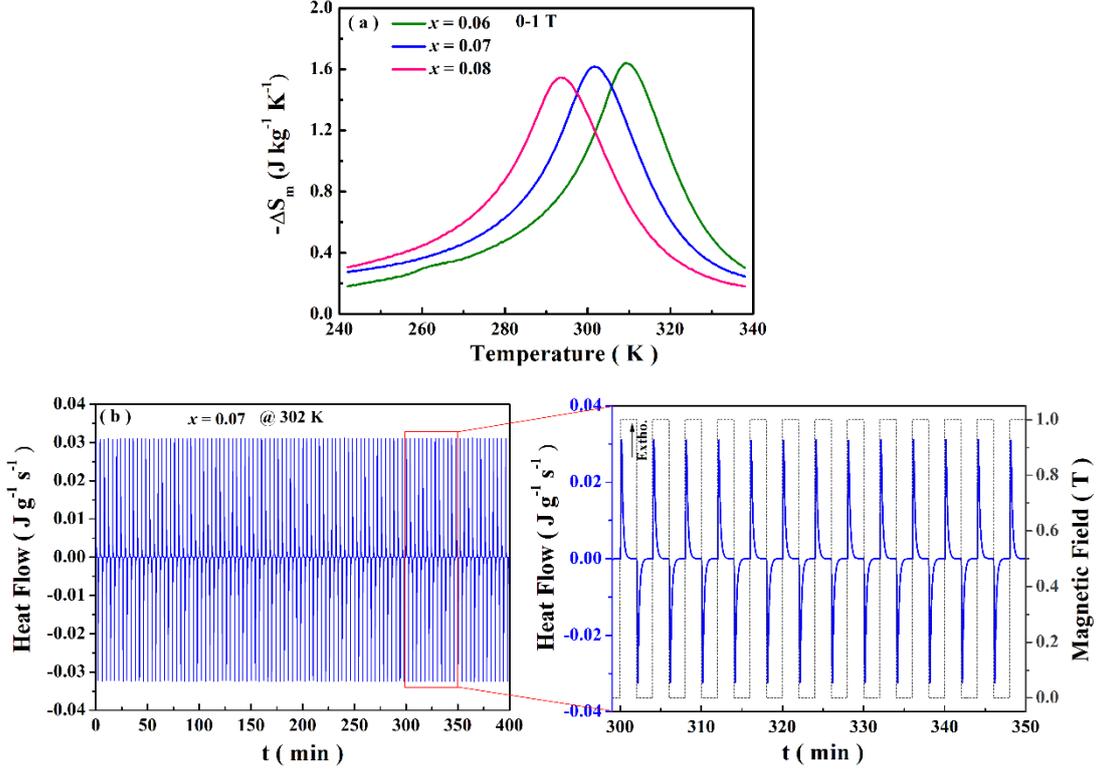

Fig. 6 (a) Temperature dependence of magnetic entropy change for a field change of 0-1 T for the MnCo$_{1-x}$Cu$_x$Ge ($x$ = 0.06, 0.07 and 0.08) alloys. (b) Heat flow for 100 cyclic magnetic fields of 0-1 T for the sample with $x$ = 0.07 at 302 K.

## 4. Discussion

In the literature $\Delta T_{hys}$ is widely adopted to identify the order of thermomagnetic phase transition. However, due to a possible temperature lag, one often observes hysteresis phenomena in calorimetric and magnetic measurements. Specifically, there exists a $\Delta T_{hys}$ of about 1 K in the sample with $x$ = 0.07 from the $M$-$T$ curves in Fig. 4b, which makes it ambiguous to determine the order of the phase transitions. Recently, Law *et al.* proposed a simple method to solve this problem quantitatively by calculating the exponent $n$ from the field dependence of $\Delta S_m$ [39]:

$$n(T,B) = \frac{d\ln(\Delta S_m)}{d\ln B} \quad (1)$$

Fig. 7a shows a 3D plot of -$\Delta S_m$ as a function of temperature and magnetic field using Maxwell relation on the basis of the isothermal $M$-$B$ curves (shown in Fig.S1 in the

Supplementary Materials):

$$\Delta S_m = \int_0^B \left(\frac{\partial M}{\partial T}\right)_B dB \qquad (2)$$

The values are in line with that obtained by DSC in field. According to the entropy data, the *n* is calculated as shown in Fig. 7b. It is evident that the *n* varies between 0.6 and 1.9 in the temperature range of the magnetic transition, which is below the limit of 2 for the crossover between first- and second-order phase transitions [39]. This indicates the second-order character of the magnetic transition for orthorhombic phase.

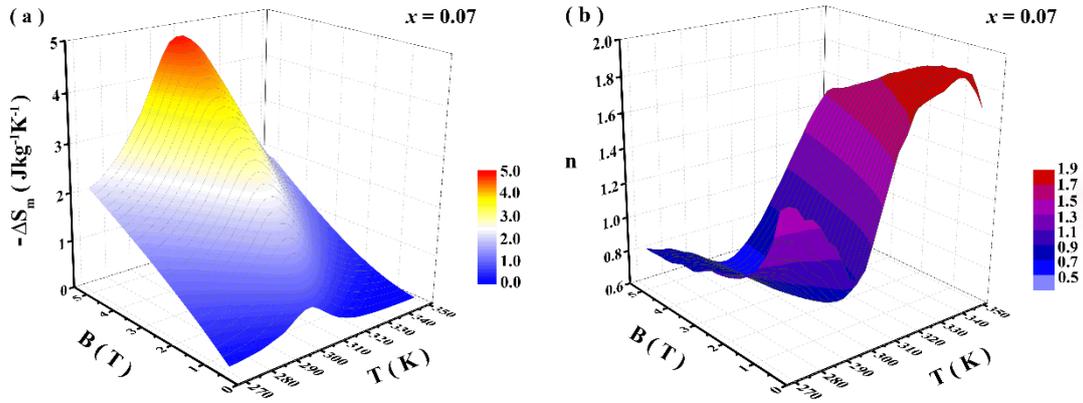

Fig. 7. 3D plot of the field and temperature dependence for (a) the magnetic entropy change and (b) exponent *n* during the magnetic transition in the sample with *x* = 0.07

In this study, a large reversible low-field MCE is achieved by stabilizing the structural transition at high temperature and decreasing $T_C^o$ to room temperature. As reported in the literature, the stoichiometric MnCoGe alloy is very sensitive to the introduction of substitutional elements, resulting in a rapid decrease in $T_t$ by increasing substitution concentrations [7,12,16,40-45]. In contrast, $T_t$ exhibits a non-monotonic change by Cu doping for Co in this work. To get insight into the origin of this experimental result, the ELF of a supercell with the hexagonal structure shown in Fig.8a is calculated for composition MnCoGe, MnCo$_{0.9375}$Cu$_{0.0625}$Ge and MnCo$_{0.75}$Cu$_{0.25}$Ge (denoted as Cu0, Cu1 and Cu4), in order to evaluate the evolution of chemical bonds [14,46]. As shown in Fig. 8b, the electrons in stoichiometric MnCoGe alloy are mainly localized around the Ge atoms, especially between the nearest-neighbor Co and Ge. The maximum ELF value is about 0.55, which gives rise to two important conclusions. Firstly, the formed covalent-like bonding between Co-Ge as the framework supports the stabilization of hexagonal phase. Secondly, the electron pairing as a result of the bonding reduces the magnetic moments in both Co and Ge atoms [47]. When Cu is introduced into the Co site, less electrons are localized between Cu-Ge atoms, while the electrons between the original Co-Ge atoms show a stronger localized character, as shown in Fig. 8c and 8d. In Fig.8e, the specific ELF values between nearest-neighbor Co/Cu and Ge atoms reflect this behavior more clearly. Consequently, the strength of the covalent bonding from the *p-d* hybridization is weakened between Cu-Ge and is enhanced between Co-Ge. The competition of both leads to the non-monotonic behavior of $T_t$.

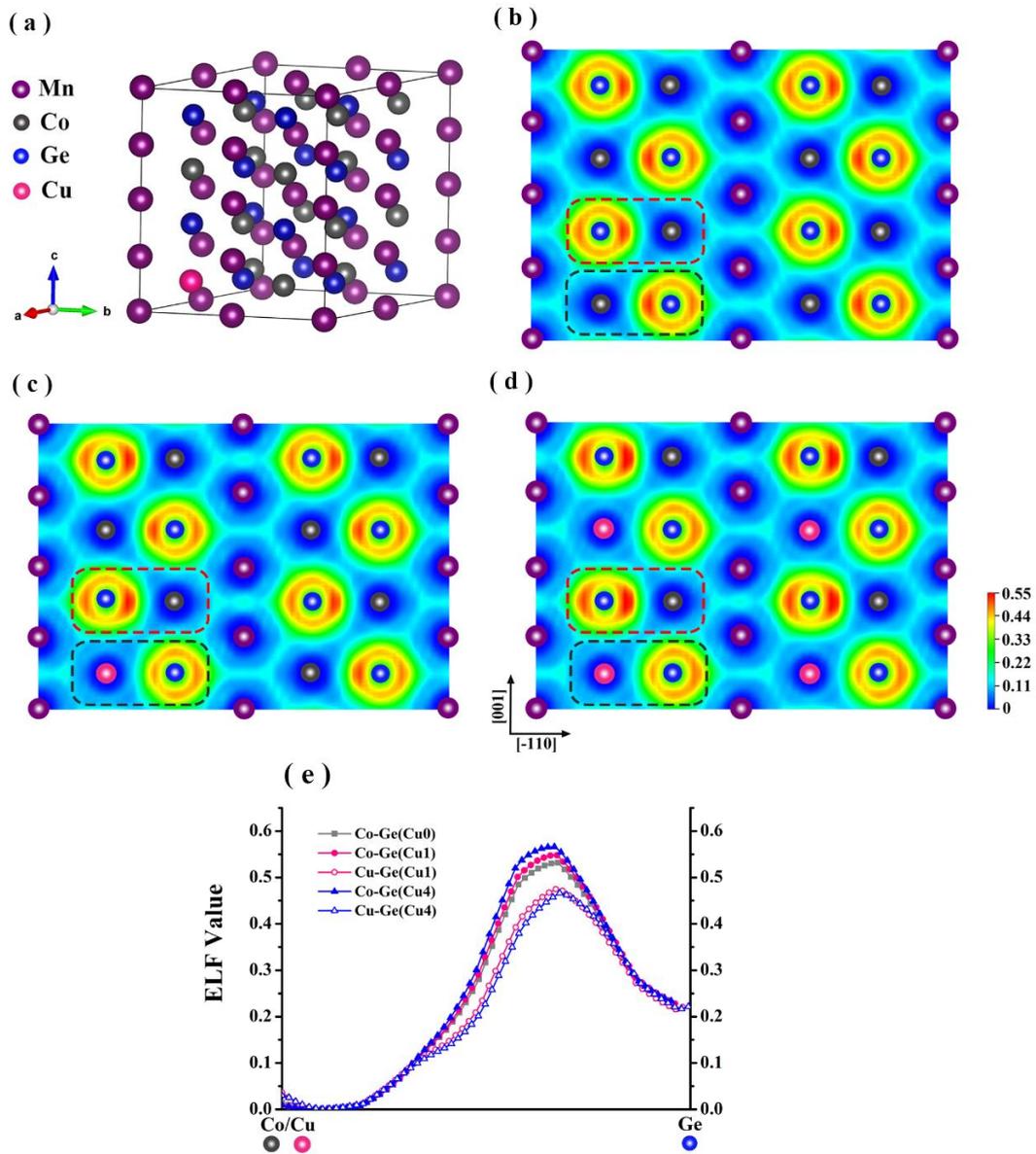

Fig. 8. (a) Crystalline structure of the MnCo$_{0.9375}$Cu$_{0.0625}$Ge supercell for the hexagonal phase. The ELF of planes (110) in (b) MnCoGe, (c) MnCo$_{0.9375}$Cu$_{0.0625}$Ge and (d) MnCo$_{0.75}$Cu$_{0.25}$Ge. (e) Variation of ELF value between nearest-neighbor Co/Cu and Ge atoms.

The exchange coupling constants $J$ between different atoms in the MnCo$_{1-x}$Cu$_x$Ge alloys with the orthorhombic structure are calculated in Fig. 9a. In stoichiometric MnCoGe the Mn-Mn and Mn-Co exchange interactions are positive and strong. This is reasonable as the magnetic moment is large for Mn atoms (3 $\mu_B$) and small for Co atoms (0.8 $\mu_B$) [48]. Considering the other interactions, the strength is weak and negligible. Notably, the Co-Co are found to show an antiferromagnetic coupling. When non-magnetic Cu is introduced into Co site, the strength of the Mn-Mn interaction slightly increases, while the Mn-Co interaction gradually deceases. Mn-Cu exhibits a weak ferromagnetic coupling and with increasing Cu content, the statistical weight of this exchange interaction increases. Based on the strength of $J$, the Curie temperature is estimated using the mean-field theory by the following equation [49,50]:

$$T_C = \frac{2}{3k_B} \cdot J_{max} \qquad (3)$$

where $J_{max}$ is the largest eigenvalue of the matrix of the exchange coupling between the atoms. This matrix has the following form:

$$\begin{pmatrix} \sum J_{Mn-Mn} & \sum J_{Mn-Co/Cu} & \sum J_{Mn-Ge} \\ \sum J_{Co/Cu-Mn} & \sum J_{Co/Cu-Co/Cu} & \sum J_{Co/Cu-Ge} \\ \sum J_{Ge-Mn} & \sum J_{Ge-Co/Cu} & \sum J_{Ge-Ge} \end{pmatrix} \qquad (4)$$

As shown in Fig. 9b, the estimated $T_C^O$ from the DFT calculations decreases monotonously with increasing Cu contents, which is consistent with the experimental behavior. In addition, the calculated $M_s$ is in good agreement with the value from $M$-$B$ curves in Fig. 4a. Evidently, the calculated $T_C^O$ is overestimated, which is widely reported for the mean-field method [51,52]. Therefore, the DFT calculations suggest that both the weak interactions between the Mn-Cu atoms and the decrease of Mn-Co interactions are responsible for the reduction of $T_C^O$.

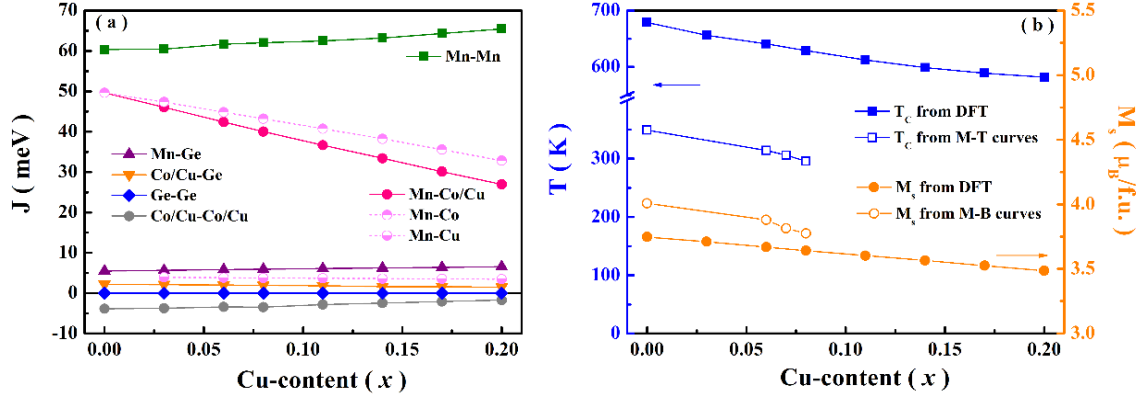

Fig. 9. (a) Calculated exchange coupling constants between the atoms in MnCo$_{1-x}$Cu$_x$Ge alloys as a function of the Cu content. (b) The estimated Curie temperature and the saturation magnetization from DFT calculations and from magnetic measurements in MnCo$_{1-x}$Cu$_x$Ge alloys, respectively.

## 5. Conclusions

In summary, we propose a new approach to design a reversible low-field magnetocaloric effect at room temperature in the MnMX family. The introduction of Cu into Co site of MnCoGe results in the stabilization of the martensitic transition at high temperatures and the decrease of $T_C^O$ in the MnCoGe alloy. Thus, a large reversible $\Delta T_{ad}$ of about 1 K under a field change of 0-1 T is achieved during the second-order magnetic transition of the orthorhombic phase. Moreover, DFT calculations are carried out to investigate the physical origin of the experimental results. The ELF analysis reveals that the strength of the covalent bonding between Cu-Ge is weakened and is enhanced between Co-Ge, which leads to a non-monotonic behavior of $T_t$. The reduction in $T_C^O$ is ascribed to a weak coupling between Mn-Cu atoms and a decrease in Mn-Co interactions. Our work deepens the understanding of the MnCoGe system and further develops its magnetocaloric performance towards practical applications.

**Acknowledgements**

The authors would like to thank A. J. E. Lefering, B. Zwart and K. Goubitz for their technical help. This work was sponsored by the National Natural Science Foundation of China (Grant Nos: 51601092, 51571121, and 11604148) and the NWO in the Domain Applied and Engineering Sciences (AES) Programme. J. Liu gratefully acknowledges financial support from the China Scholarship Council.